# Automated RNA structure prediction uncovers a missing link in double glycine riboswitches

Wipapat Kladwang[1], Fang-Chieh Chou[2], Rhiju Das[1,3,*]

Departments of Biochemistry[1], Chemistry[2], and Physics[3], Stanford University, Stanford, CA 94305, US

**ABSTRACT:** The tertiary structures of functional RNA molecules remain difficult to decipher. A new generation of automated RNA structure prediction methods may help address these challenges but have not yet been experimentally validated. Here we apply four prediction tools to a remarkable class of double glycine riboswitches that exhibit ligand-binding cooperativity. A novel method (BPPalign), RMdetect, JAR3D, and Rosetta 3D modeling give consistent predictions for a new stem P0 and kink-turn motif. These elements structure the linker between the RNAs' double aptamers. Chemical mapping on the *F. nucleatum* riboswitch with SHAPE, DMS, and CMCT probing, mutate-and-map studies, and mutation/rescue experiments all provide strong evidence for the structured linker. Under solution conditions that separate two glycine binding transitions, disrupting this helix-junction-helix structure gives 120-fold and 6- to 30-fold poorer association constants for the two transitions, corresponding to an overall energetic impact of $4.3 \pm 0.5$ kcal/mol. Prior biochemical and crystallographic studies from several labs did not include this critical element due to over-truncation of the RNA. We argue that several further undiscovered elements are likely to exist in the flanking regions of this and other RNA switches, and automated prediction tools can now play a powerful role in their detection and dissection.

Non-coding RNA sequences play critical roles in cellular biochemistry and genetic regulation, and their number is growing rapidly.[1,2] Many of these RNAs' behaviors are intimately tied to their three-dimensional conformations[3], but determining these structures has challenged both experimentalists and modelers, especially with regards to tertiary interactions mediated by non-Watson-Crick base pairs. Recent years have seen the development of automated algorithms for detecting and modeling RNA tertiary structure, especially modular recurrent motifs.[4-8] However, the predictive power of these methods has yet to be demonstrated through rigorous experiments. Here we report the application and chemical validation of RNA structure prediction to discover a previously missed tertiary element in a paradigmatic class of natural RNA riboswitches.

The cooperative binding of small molecules is a fundamental feature of functional biopolymers that has been studied in numerous model systems[9], most of which were protein-based until 2004. In that year, Breaker and co-workers reported ligand-binding cooperativity in an RNA riboswitch that binds two glycine molecules[10]. This discovery has inspired many biophysical studies[10-14], culminating in the recent publication of crystallographic models of double aptamers[15,16], but a predictive model for ligand-binding cooperativity has remained elusive. Prior investigations have mostly focused on constructs pared down to minimal sequences required for glycine binding. We therefore sought to explore outer flanking sequences that have so far remained uncharacterized. We started by applying four recently developed RNA structure modeling tools and found remarkably consistent predictions for a new motif.

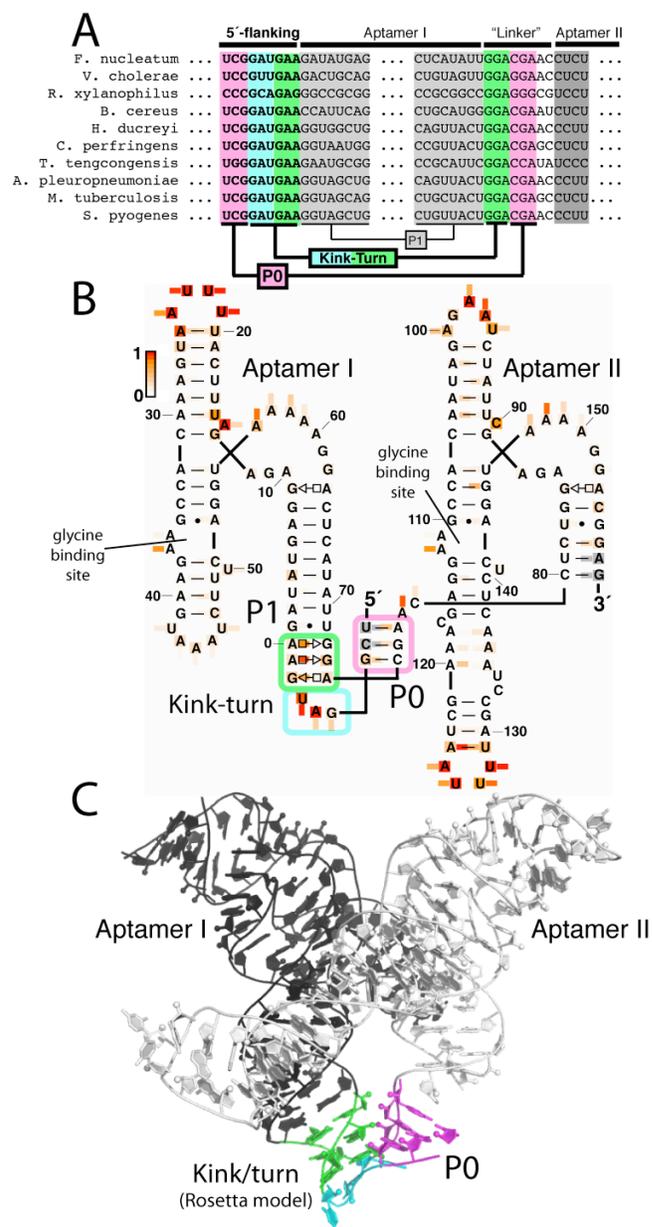

FIGURE 1. A new stem P0 and kink-turn tertiary element in double glycine riboswitches on a sequence alignment (A), on the secondary structure of the *F. nucleatum* riboswitch (B), and built into this RNA's 3D crystallographic model (C). Coloring in (B) shows normalized chemical mapping data: SHAPE (on letters), DMS (rectangles at A/C), and CMCT (rectangles at G/U).

As input into modeling, we extracted aligned sequences of 360 double glycine riboswitches, collating single-aptamer entries in RFAM[17] and extending these alignments into 5′ and 3′ flanking regions by 100 nucleotides. Inspired by consensus approaches for protein fold recognition[18], our new BPPalign tool (SI Methods) searches for novel stems by averaging base-pair probability calculations for Watson-Crick secondary structure across homologs. In addition to new candidate elements in the 3′ region (SI Fig. S1) containing potential 'expression platforms'[7,10,19], we found signals involving nine nucleotides preceding the conventional start of the riboswitch. For 160 sequences, including the widely studied riboswitches from *Vibrio cholerae*[10,11,13] and *Fusobacterium nucleatum* (FN)[12,14], a three base-pair interaction between the riboswitch inter-aptamer linker and nucleotides in the 5′ flanking sequence could occur (SI Fig S1 and Fig. 1). We call this putative stem P0. In addition, the nearby P1 stem of the first aptamer was found to potentially form an extended 3 three purine-purine pairs (129 sequences; Fig. 1) or, in some cases, three Watson-Crick pairs (41 sequences). Finally, while P0 and this extended P1 are contiguous on the 3′ strand, there is an intervening three-residue bulge in the 5′ strand. We noted that the lengths and sequences of these features matched the published consensus for the kink-turn motif[20]. This motif, originally described based on high-resolution ribosome crystallography[21], has been annotated in numerous functional RNAs[7,20] but never previously reported in glycine riboswitches.

Three independent lines of bioinformatic/computational evidence supported the presence of the kink-turn motif. First, the automated RMdetect software[7] gave a strong-confidence detection of a kink-turn (101 sequences, SI Fig. S2) when given our extended multiple sequence alignment. Second, the JAR3D server[8,22] returned a kink-turn motif when given the putative P0/P1 sequences (SI Fig. S3). Third, we applied Rosetta/FARFAR 3D modeling[6] to replace the linker in the FN riboswitch crystallographic model[16] with a kink-turn motif forming a continuous helical interface with P1. The resulting model (Fig. 1C) demonstrated that a kink-turn can bridge the riboswitch aptamers with reasonable geometry and no clashes. Furthermore, we carried out *de novo* modeling of the entire linker and added 5′ strand; encouragingly, the lowest-energy models exhibited kink-turn conformations (SI Fig. S4). Finally, an unexpected concordance was observed: the linker backbone within the modeled kink-turn approximately followed the linker path in the deposited crystallographic electron density and coordinates[16] (4.4 Å C4′ RMSD), despite the absence of any pairing partners in the crystallized molecule (SI. Fig. S5).

The computational approaches applied above gave consistent predictions, but these tools are largely untested, can give false positives (SI Figs. 1 and 2), and do not provide information on the energetic significance of the putative element. Therefore, we carried out experiments to confirm and further characterize the kink-turn motif using mutate-and-map and mutate/rescue trials, with quantitative readouts from nucleotide-resolution chemical mapping. As a model RNA, we chose the smallest known glycine riboswitch, the *FN* system, which has been extensively studied by mutation, chemical mapping, and crystallography, albeit in truncated form. We focused on a sequence with a 9-nucleotide natural extension restored to the 5′ end compared to the prior construct (Fig. 1B; SI Methods). We called this sequence FN-KTtest. High throughput measurements of dimethyl sulfate (DMS) and carbodiimide (CMCT) modification reported on the chemical accessibilities of Watson-Crick nucleobase edges for A/C and G/U, respectively[23]. Nucleotides outside the linker region served as controls in this analysis; we confirmed that their reactivities in 10 mM MgCl₂, 50 mM Na-HEPES, pH 8.0, and 10 mM glycine correlated with the burial of bases in the glycine-bound *FN* crystallographic model[16] (Fig. 1B). Within the linker region and within the added natural 5′ flanking sequence, the DMS and CMCT reactivities were consistent with the 3D model of the kink-turn (protections of nts −2, −4 to −6, and 72 to 77; and exposure of 0, −1, and −3; Figs. 1B and C). Further analysis, based on 2′-OH acylation with N-methylisatoic acid (SHAPE chemistry[24]), gave protections of linker nucleotides 72-79 (Fig. 1B), as would be expected for a structured element; these nucleotides gave high SHAPE reactivities in previous studies without the 5′ flanking sequence.[12,14]

More stringent tests of the predicted kink-turn structure were achieved with mutate-and-map[14,23,25] experiments, shown in Fig. 2. We first disrupted the 5′ end of putative P0 stem ( $\frac{5'-UCG-3'}{3'-AGC-5'}$ to $\frac{5'-AGC-3'}{3'-AGC-5'}$ at nts −8 to −6 and 75 to 77; called MutA), expecting the reactivity of the mutated segment and its base-pairing partner to increase. Indeed, the SHAPE reactivities in both P0 strands as well as in putative P1 purine/purine pairs (nts −2 to 0 and 72−74), increased to the level seen in unpaired loops (see, e.g., nts 20−25 and 96−100) and to the level seen in the prior construct without the 5′ flanking sequence[26]. SHAPE reactivity in other regions of the glycine riboswitch did not change significantly. We also mutated the 3′ end of P0 ( $\frac{5'-UCG-3'}{3'-UCG-5'}$ , called MutB), and observed disruptions in the same regions. Finally, we sought to 'rescue' these structural disruptions by implementing both sets of mutations ( $\frac{5'-AGC-3'}{3'-UCG-5'}$ , called MutAB). The resulting variant indeed restored SHAPE reactivities to levels observed for the wild type FN-KTtest RNA. DMS and CMCT chemical mapping data gave analogous results (Fig. 2). We conclude that there is strong biochemical evidence for the P0 stem and P1 purine pairs, and, combined with the computational and chemical mapping data above, strong evidence for the kink-turn motif in the FN glycine riboswitch. Analogous measurements in the absence of glycine (Fig. 2) showed that this motif is also formed in the glycine-free state of the RNA.

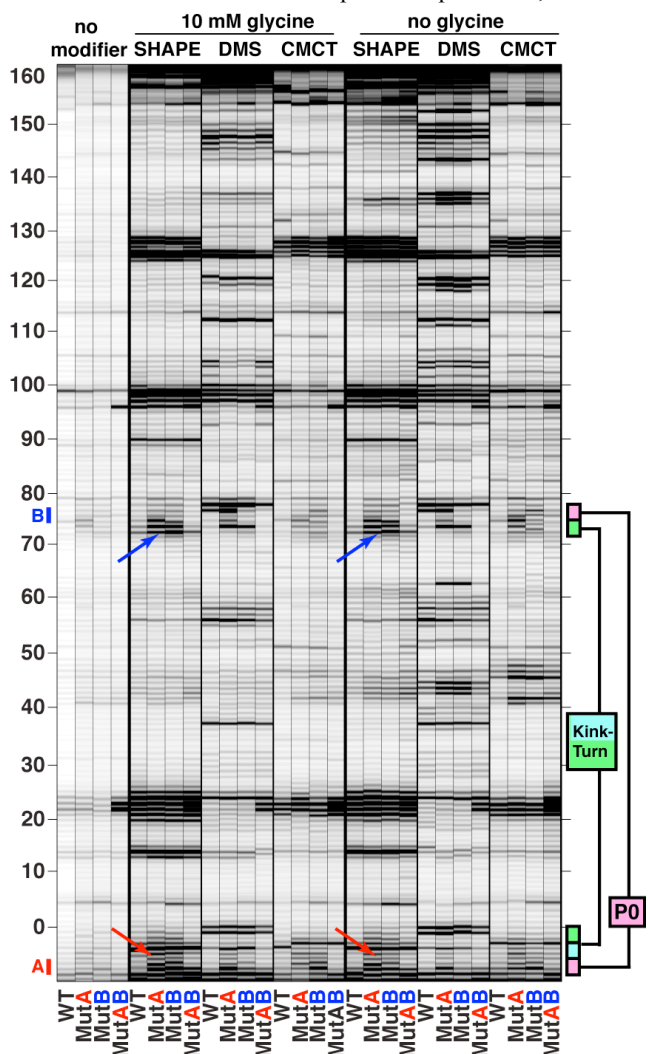

FIGURE 2. Evidence for the predicted kink-turn tertiary element and P0 stem from disruption of chemical reactivities by mutation (MutA and MutB) and restoration (MutAB) of P0 Watson-Crick pairs. Fluorescence from aligned capillary electrophoretic traces from chemical mapping are shown. Arrows mark SHAPE effects discussed in text.

We then sought to determine the energetic importance of the kink-turn motif for riboswitch glycine binding. We monitored glycine-induced conformational changes with DMS chemical mapping (which gave stronger reactivity differences than CMCT or SHAPE). Quantitative thermodynamic comparisons require that ligand-binding events monitored for different molecule variants correspond to transitions between analogous states. In the experiments above, the wild type FN-KTtest and rescued double mutant MutAB appeared partially folded even in glycine-free conditions, giving DMS protections at nts 43−45, 63, and 135−137 compared to single-strand mutants MutA and MutB. We therefore lowered the MgCl₂ concentration from 10 mM to 0.5 mM; under these conditions all variants gave indistinguishable chemical reactivities (outside the kink-turn linker) in their glycine-free states. For all constructs, two glycine-dependent transitions could be resolved (Fig. 3), permitting the characterization of the kink-turn's energetic effect on individual glycine-binding events. The equilibrium is described in terms of glycine-free, single-glycine-bound, and two-glycine-bound states:

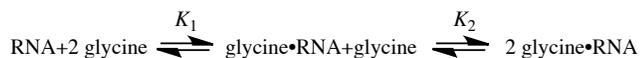

$$\text{RNA+2 glycine} \xrightleftharpoons{K_1} \text{glycine•RNA+glycine} \xrightleftharpoons{K_2} \text{2 glycine•RNA}$$

We inferred equilibrium constants $K_1$ and $K_2$ and chemical reactivities for each state via a likelihood-based analysis. Further discussion of the fitted reactivities, interpretation of $K_1$ and $K_2$, error estimation from replicates, and comparison to Hill-type fits is given in SI Methods, SI Text, and SI Figs. S6 and S7.

As a baseline, the wild type FN-KTtest construct gave $K_1 = 9 \pm 0.5$ μM and $K_2 = 1.8 \pm 0.4$ mM (Fig. 3; see SI Fig. 6 for full data). The mutants MutA and MutB then provided independent tests of the energetic impact of the kink-turn element. They required substantially higher glycine to undergo conformational change, with $K_1$ increased by 120-fold and $K_2$ increased by 6-fold to 30-fold ($K_1 = 1.1 \pm 0.2$ mM and $K_2 = 50 \pm 20$ mM for MutA; $K_1 = 1.1 \pm 0.2$ mM and $K_2 = 10 \pm 3$ mM for MutB). The disruption of the newly discovered linker thus results in free energy perturbations of $2.8 \pm 0.1$ kcal/mol and an additional $1.5 \pm 0.5$ kcal/mol for the two glycine binding events, respectively. As expected, the double mutant MutAB restored the equilibrium constants to $K_1 = 10 \pm 1$ μM and $K_2 = 1.3 \pm 0.5$ mM, indistinguishable from the wild type equilibria within experimental errors.

These measurements confirmed unambiguously that this kink-turn motif – although missed in seven years of intense biochemical studies – has a large energetic impact on glycine riboswitch behavior. The inclusion of this motif in future work may lead to more precise chemogenetic data[12], more easily interpretable x-ray scattering profiles[11], and to better diffracting crystals[15,16] with and without glycine. Interestingly, the kink-turn enhances the affinity but not the thermodynamic cooperativity of glycine binding (as would be reflected in increased $K_1/K_2$). The $FN$ glycine riboswitch may not have evolved to act cooperatively; or it may require different solution conditions, other molecular partners (e.g., kink-turn-binding proteins[27]), or a kinetic mechanism to exhibit ligand-binding cooperativity[28,29]. Given our results herein, we favor another hypothesis: there are secondary and tertiary motifs even further out in the RNA's flanking regions that modulate the riboswitch's behavior. Some candidate interactions are listed in SI Text. There are certainly precedents for flanking sequences playing unexpected roles in functional RNAs (e.g. refs[30-32]). We expect automated RNA structure prediction to be a powerful tool for uncovering missing pieces of the double glycine riboswitch and of other functional RNAs that remain mysterious.

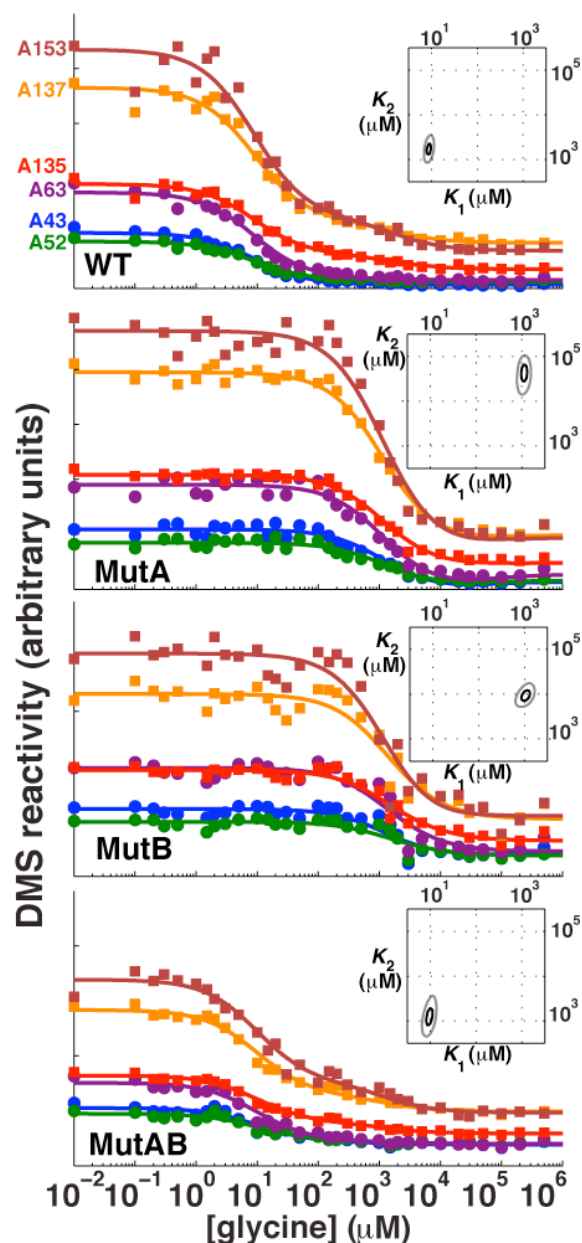

FIGURE 3. Thermodynamic analysis of glycine binding to riboswitches with (WT, MutAB) and without (MutA, MutB) the kink-turn. Symbols show DMS reactivities for six representative nucleotides (see top panel); colored lines give calculated reactivities for a model with two glycine-binding events. See SI Fig. S6 for complete data. Insets: Log-likelihood contours for $K_1$ and $K_2$ at 2 (black) and 10 (gray) units below maximum likelihood value.

## ASSOCIATED CONTENT

**Supporting Information**. Methods, supporting text, and seven figures describing structure prediction, chemical mapping, and thermodynamic fitting.

## AUTHOR INFORMATION

### Corresponding Author

* Phone: (650) 723-5976. Fax: (650) 723-6783. E-mail: rhi-ju@stanford.edu.

### Author Contributions

W.K. carried out experiments; F.C. carried out Rosetta modeling; R.D. designed research, carried out modeling and analysis, and wrote the manuscript. All authors have given approval to the final version of the manuscript.

### Funding Sources

Burroughs-Wellcome Career Award at the Scientific Interface (R.D.), and Study Abroad Scholarship of Taiwan Government (F.C.).

## ACKNOWLEDGMENT

We thank P. Sripakdeevong for help in Rosetta modeling; C. VanLang and J. Sales-Lee for discussions; and A. Petrov, C. Zirbel, and N. Leontis for early access to the JAR3D server. Rosetta computations were carried out on the BioX$^2$ cluster (NSF CNS-0619926).

## ABBREVIATIONS

FN, Fusobacterium nucleatum; nts, nucleotides; DMS, dimethyl sulfate; CMCT, 1-cyclohexyl-3-(2-morpholinoethyl) carbodiimide metho-p-toluenesulfonate; SHAPE, selective hydroxyl acylation analyzed by primer extension.

**Table of Contents Figure**

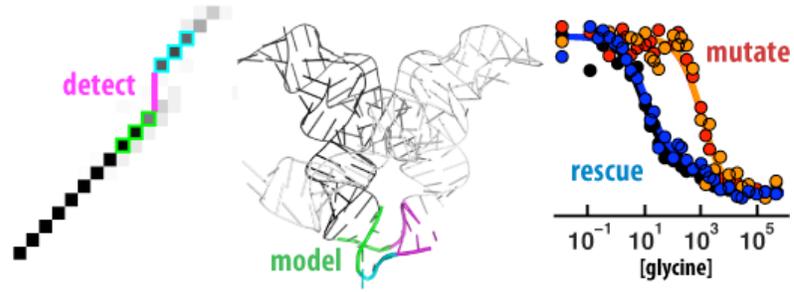

Supporting Information for **"Automated RNA structure prediction uncovers a missing link in double glycine riboswitches"**

Wipapat Kladwang[1], Fang-Chieh Chou[2], Rhiju Das[1,3,*]

Departments of Biochemistry[1], Chemistry[2], and Physics[3], Stanford University, Stanford, CA 94305, US

This documents contains Supporting Methods, Supporting Text, Supporting Referencs and Supporting Figures S1–S7.

# Supporting Methods

## *Preparation of sequence alignments*

The Rfam database (v10.0, January 2010) contained 3368 sequences in the glycine riboswitch entry. Double riboswitches were assembled by searching for sequences drawn from the same genomic source in which the end of one domain was within 9 nucleotides of the start of a second domain. 86% of the Rfam sequences passed this criterion, giving 1453 double riboswitches. We further filtered this alignment to remove marine shotgun metagenome sequences (which appeared highly redundant for this domain), leaving 360 sequences. Flanking sequences (100 nucleotides 5′ and 100 nucleotides 3′ of the Rfam seed boundaries) and any missing inter-domain linker sequences were filled in via *efetch* queries to the Entrez database of the form

```
http://eutils.ncbi.nlm.nih.gov/entrez/eutils/efetch.fcgi?db=nucleotide
```

All steps were carried out using automated python scripts, available with the *BPPalign* tool, described next

## *The* BPPalign *method*

The *BPPalign* method was inspired by approaches in protein modeling that seek to improve the accuracy of structure predictions by evaluating consensus across several fold recognition algorithms[1] or different homologs[2]. The protocol involved two steps. First, all sequences in the alignment above were folded in the *partition* executable in *RNAstructure*[3] (version 5.3), with the base pairing probability *bpp* matrix saved to disk. Then, the *bpp* data were averaged across all homologs, based on alignment to one reference sequence (here, the *F. nucleatum* riboswitch sequence from NCBI accession number AE009951.2). This aligning and averaging were carried out in MATLAB (R2010b), using a script *make_bpp_plot.m*. All sequence handling scripts are available through the SimTK website under the project BPPalign (https://simtk.org/home/bpp_align).

## *RMdetect*

RMdetect[4] (version 0.0.3) was applied to the same multiple sequence alignment as the BPPalign tool above. The command lines used were:

```
rmdetect.py  new_align_no_metagenome.stk > rmdetect_result.txt

rmcluster.py --min-occur=0.05 --min-score=9.0 --min-bpp=0.01
  --min-mi=0.001 --out-dir=./ --fig < rmdetect_result.txt >
  rmdetect_cluster.txt

rmout.py < cluster_004_KT_1.0.res  > cluster_004_KT_1.0_examples.txt
```

We searched for modules  CL_1.0 (C-loop), GB_1.0 (G-bulge), KT_1.0 (kink-turn), TGA_1.0 (tandem G/A). In separate runs, we verified the recovery of the newer ARICH modules; these results are not shown since this module's definition was actually based on the glycine riboswitch Rfam alignment. We note that previous work applying RMdetect to the glycine riboswitch did not find the kink-turn reported herein due to the use of truncated single-domain alignments from Rfam.

*JAR3D*

JAR3D aligns given RNA sequences to hybrid stochastic-context-free-grammar/Markov-random-field models[5-7]. We used a beta version of the JAR3D web server. The server was given the sequence drawn from the *FN* riboswitch, GGAUGAAG*UGGAC, where '*' separates the strands. Nine motifs achieved over 90[th] percentile score for this sequence, and all were kink-turn motifs or close variants; the top hit (Group_311; percentile 100%) is shown in SI Fig. S3. Similar results were achieved when the entire alignment of sequence homologs in these segments were given to JAR3D.

*Rosetta grafting*

First, the sequence of the proposed kink-turn in the glycine riboswitch was compared with the ones of known kink turns[8]. We found that the *H. marismortui* 23S ribosomal RNA KT-46 motif has the most similar sequence to the proposed kink-turn sequence in the *FN* glycine riboswitch: they both start with two G-C base pairs, followed by a 3-nucleotide bulge, then 4 non-Watson-Crick (WC) base pairs of the same sequence at the end:

```
KT in FN riboswitch (3P49)
      GAU
  5´ CG   GAAG    -7 - 1
  3´ GC---AGGU    76 - 71
```

```
KT-46 from H. marismortui 23S rRNA (3CC2)
      AUG
  5´ GG   GAAG    1311 - 1319
  3´ CC---AGGU    1343 - 1338
```

Therefore, we took nucleotides 1310–1323/1334–1344, which included the KT-46 motif, from the PDB coordinates 3CC2.[9] The WC base-pairing region (nts 1320–1323/1334–1337) was aligned to the corresponding region (nts 2–5/67–70) in the glycine riboswitch model 3P49[10]; and we replaced the proposed kink-turn region (nts 0–2/70–77) in the glycine riboswitch with the KT-46 model (nts 1310–1320/1337–1344), with the nucleobase mutated in Rosetta[11,12]. The junction nucleotides (nts 2 and 70) were minimized with variable bond length and bond angle, as in ref[11]. The 2-nucleotide loop connecting the kink turn and the second aptamer (nts. 78–79) was modeled using FARFAR loop modeling[12], generating 100 models each with 5000 fragment insertions. The lowest energy model is shown in Fig. 1C.

*Rosetta de novo structure prediction.*

For fully *de novo* modeling, we removed nucleotides 0, 1 and 71 in the P1 stem, and 72–79 in the linker from the crystallographic model 3P49[10]. We then rebuilt strands −7 to 1 and 71–79 with Rosetta Fragment Assembly of RNA with Full Atom Refinement (FARFAR[11]). 50,000 models were generated with 10,000 fragment insertions each, and the lowest energy 1000 models were clustered with RMSD threshold 2.0 Å. Four of five lowest energy cluster centers recovered the P0 stem, at least 2 of 3 sheared purine pairs, and a sharp bend at the three-nucleotide bulge (SI Fig. S4). One of the cluster centers recovered all three purine pairs and a complete kink-turn conformation; this model gave an all-atom RMSD of 3.4 Å over the rebuilt nucleotides from the 'graft' model above (SI Fig. S4).

*RNA preparation and high-throughput chemical mapping*
The full FN-KT RNA sequence included 5′ and 3′ flanking sequences to permit interpretation of chemical mapping data all the way to the beginning and end of the riboswitch domains and to permit facile primer binding; the sequence was:

```
ggaaauaaUCGGAUGAAGAUAUGAGGAGAGAUUUCAUUUUAAUGAAACACCGAAGAAGUAAAUCUUUCAGG
UAAAAAGGACUCAUAUUGGACGAACCUCUGGAGAGCUUAUCUAAGAGAUAACACCGAAGGAGCAAAGCUAA
UUUUAGCCUAAACUCUCAGGUAAAAGGACGGAGaaaacaaaacaaagaaacaacaacaacaac
```

The primer binding site is underlined. Sites that were mutated to their complements in MutA, MutB, and MutAB are given in boldface. Flanking sequences are in lowercase. The added flanking sequences were checked in *RNAstructure*[13] and *Viennafold*[14] to give no predicted base pairing with the glycine-binding domain; chemical mapping data for parts of the flanking sequences were observable and gave high reactivity, as expected. Further, agreement of chemical reactivity in the glycine-binding domain (outside the predicted kink-turn) to prior measurements with different flanking sequences confirmed the lack of interaction of flanking sequences with the glycine-binding domain.

Preparation and purification of RNAs, chemical mapping with DMS, CMCT, and NMIA (SHAPE) modification, and the readout of modifications by reverse transcription and capillary electrophoresis were carried out as previously described[11,15]. Briefly, DNA templates for the RNAs were prepared by PCR assembly, with a 20-bp promoter TTCTAATACGACTCACTATA for T7 RNA polymerase, and purified with AMPure magnetic beads (Agencourt, Beckman Coulter). To remove a mispriming product, these DNA samples were further purified by agarose gel electrophoresis, band excision, and purification in Qiaquick microcentrifugation columns following kit instructions. Then, 40 μL *in vitro* RNA transcriptions were purified with MagMax magnetic beads (Ambion). RNAs (at 60 nM concentration) were incubated in the desired solution conditions (e.g., 10 mM $MgCl_2$, 50 mM Na-HEPES, pH 8.0, and 10 mM glycine) for 10 minutes before addition of chemical modification reagent (final concentrations: 0.125% DMS, 2.6 mg/mL CMCT, or 3.0 mg/mL N-methyl isatoic acid) and incubation times of 15 minutes (DMS, CMCT) or 30 minutes (NMIA) at 24 °C. We were able to achieve final glycine concentrations of 500 mM before reaching solubility limits; we note here glycine concentrations higher than 10 mM appear to quench SHAPE measurements (NMIA), but DMS and CMCT reactions were robust up to 500 mM glycine (C. VanLang, RD, unpublished data).

Chemical modification reactions were quenched with β-mercaptoethanol (DMS) or lowered pH (Na-MES, pH 6.0); purified by binding to a fluorescent (5′-rhodamine-green-labeled) DNA primer pre-bound to poly-dT beads; reverse transcribed with SuperScript III enzyme; and purified and desalted before co-loading on ABI 3100 or 3730 capillary sequencers with Texas-red-labeled reference ladders. All data were analyzed in the HiTRACE software[16]. For Fig. 1B (estimates of intrinsic chemical reactivities), data were background-subtracted and corrected for attenuation of reverse transcription products as in ref.[17]. For glycine titration analyses, these extra processing steps were not used in order to avoid introducing additional noise.

*Analysis of titrations: thermodynamic formalism and likelihood-based fitting*

Our system involves the binding of at most two ligands and can be formally described by the chemical equilibrium

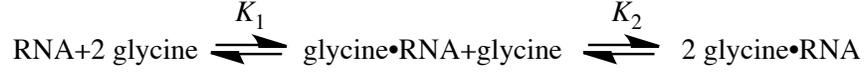

$$\text{RNA+2 glycine} \overset{K_1}{\rightleftharpoons} \text{glycine•RNA+glycine} \overset{K_2}{\rightleftharpoons} \text{2 glycine•RNA}$$

or, equivalently, the partition function:

$$Z = 1 + \frac{x}{K_1} + \left(\frac{x}{K_1}\right)\left(\frac{x}{K_2}\right),$$ (1)

where $x$ is the ligand concentration[18]. The terms partition all RNA conformations without ligand bound into one state **0**, all conformations with one ligand bound into state **1**, and all double-ligand-bound conformations into the state **2**. Systems with high binding cooperativity correspond to cases with low detectable population of state 1 at all $x$, as happens with $K_1 > K_2$. It is important to note that state **1** may be an admixture of multiple structures (e.g., with different sites bound to the single ligand) or a single structure; a glycine titration on a single RNA cannot distinguish between the two, but comparison of RNA mutants can discriminate models (see caption of SI Fig. S5).

The fraction of conformations within the three states are then:

$$f_0 = \frac{1}{Z}$$

$$f_1 = \frac{\dfrac{x}{K_1}}{Z}$$ (2)

$$f_2 = \frac{\left(\dfrac{x}{K_1}\right)\left(\dfrac{x}{K_2}\right)}{Z}$$

The nucleotide-by-nucleotide chemical mapping data for each RNA variant can be summarized in data matrix $D_{ij}$ where $i$ indexes across the glycine concentrations and $j$ indexes across nucleotides. The predictions for these data, $D_{ij}^{pred}$ are given by the summing the chemical reactivity for each state $C_j^a$, weighted by the population fractions in each state:

$$D_{ij}^{pred} = \sum_{a=0,1,2} f_a^i C_j^a,$$ (3)

where $f_a^i$ are the population fractions of state $a$ at glycine concentration $i$. We carried out fits of $D_{ij}$ to $D_{ij}^{pred}(K_1, K_2)$ by grid search over $K_1$ and $K_2$ (each from $10^1$ to $10^5$ μM, in $10^{0.05}$ increments), obtaining analytical solutions for $C_j^a$ maximum likelihood values. The likelihood function[19] was given by:

$$L \propto \prod_{i,j} \frac{1}{\sigma_j} \exp\left[ -\frac{\left(D_{ij} - D_{ij}^{pred}\right)^2}{2\sigma_j^2} \right]$$

Here we have allowed each nucleotide $j$ to have a different error. Given $K_1$ and $K_2$ (and thus $f_a^i$), this likelihood function is optimized by setting its derivatives with respect to each parameter $C_j^a$ to zero, resulting in the following linear equation at each $j$:

$$\sum_{i,a'} f_a^i f_{a'}^i C_j^{a'} = \sum_i f_a^i D_{ij}$$

The errors are then solved as:

$$\sigma_j^2 = \frac{1}{N_i} \sum_i \left( D_{ij} - D_{ij}^{pred} \right)^2$$

The log-likelihood function then simply evaluates to $\log L = -N_i \sum \log \sigma_j + \text{constant}$.

The procedure above gave well-defined solutions for $K_1$ and $K_2$ for the wild type and MutAB glycine titration data, where two distinct transitions could be resolved visually (Fig. 3 and SI Fig. 4). For MutA and MutB data, the transitions were closer together, and the analysis above gave several distinct local likelihood optima, some involving anomalous transitions with $K_1$ or $K_2$ above the largest measured concentrations (0.5 M). To regularize the fits, we assumed that the reactivities for the states **0**, **1**, and **2** were similar to $C_j^a[\text{FIT}]$ measured for the wild type and MutAB constructs (indeed this assumption is required in order to make any thermodynamic comparison). The log-likelihood function was supplemented with a term:

$$L_{regularized} = L \exp\left[ -\beta \sum_{a,j} \frac{1}{2\sigma_j^2} \left( C_j^a - C_j^a[\text{FIT}] \right)^2 \right]$$

with β = 0.1; variation of this parameter by 2-fold did not change the results.

To determine final fitted values for $K_1$ and $K_2$ for each construct, $\log L(K_1, K_2)$ was summed across replicate measurements taken on different days (two for wild type and MutA; three for MutB and MutAB; each set involved two independent preparations of RNA). See insets to Fig. 3. Values reported in the main text are maximum likelihood values for $K_1$ and $K_2$. Errors give the bounds of the log-likelihood contour corresponding to a decrement of 2 from the likelihood value; these bounds correspond closely to twice the standard error.[19]

Free energy perturbations from mutations were derived from the Boltzmann relation and the assumed partitition function. If the wild type construct gives best-fit equilibrium constants $K_1$ and $K_2$ and a mutant gives $K_1'$ and $K_2'$, the free energy perturbations for binding the first glycine is:

$$\Delta\Delta G_{0 \to 1} = k_B T \log(K_1' / K_1)$$

And for binding the next glycine is:

$$\Delta\Delta G_{1 \to 2} = k_B T \log(K_2' / K_2)$$

The free energy perturbation for binding two glycines to the ligand-free state is the sum of the above two expressions: $\Delta\Delta G_{0 \to 2} = \Delta\Delta G_{0 \to 1} + \Delta\Delta G_{1 \to 2}$.

# Supporting Text

*Comments on the Hill equation*
In some prior work[20-22], thermodynamic measurements were fit to a Hill-like equation:

$$f_{\text{unbound}} = 1 / [1 + (x / K)^n]; \ f_{\text{bound}} = (x / K)^n / [1 + (x / K)^n]$$

This fit is useful as a phenomenological characterization in cases of unknown binding stoichiometry. However, the Hill equation does not correspond to a realistic model except in special cases where $n$ is an integer; those cases correspond to models which exhibit exactly two states, with 0 or $n$ ligands bound, and do not apply here. This issue is well-known in studies of multiple ligand binding by proteins (see, e.g., ref[23] for review of the thermodynamic formalism for hemoglobin oxygen-binding). For the present problem, using the Hill-like equation at different nucleotides or with different probes (e.g., different chemical modifications; gel mobility; x-ray scattering) gives different numerical values for $K$ and $n$. In contrast, the thermodynamic model defined above in eqs. (1)–(3) should give the same values (within experimental error) for $K_1$ and $K_2$ with different measurements. We have verified this by, e.g., repeating our analyses for DMS reactivity at just nucleotides 135–150; we arrive at indistinguishable results.

*Candidates for further glycine riboswitch interactions*

From the BPPalign output as well as manual inspection of the double glycine riboswitch alignment, we found candidates for additional elements that may be important for riboswitch function. These are under experimental investigation; we list them here to encourage their validation or falsification by other groups.

1. <u>Expression platforms.</u> Transcription terminator and translational suppressor stems 3´ of the ligand binding domains are readily identifiable (SI Fig. S1) and in several cases are mutually exclusive with P1 of the second aptamer. In the *FN* riboswich, three potential terminator hairpins with multiple subsequent U-nucleotides can be identified.

2. <u>CUCUC/GAGAG</u> The sequence CUCUC (or similar) appears at the glycine binding site in nearly all riboswitches in both aptamers I and II (e.g., nts 49–53 and nts 138–142 in the *FN* sequence). It is also noteworthy that the P1 stem of aptamer II typically contains sequence CUCU. The sequence GAGAG (or similar) appears in nearly all riboswitches (e,g, nts 9–13 and nts 86–90) in the core 3-way junction; BPPalign (SI. Fig. S1) often pairs this to the CUCU sequences. These interactions are not consistent with the glycine-bound crystallographic structure and therefore may be important in the glycine-free state. However we have not been able to detect covariation across these potentially interacting sequences; they may be conserved for other reasons including stability of the glycine-bound state.

3. <u>Upstream modulators.</u> The BPPalign analysis shows several potential elements 5´ of the domain that may disrupt P1 or P0 through alternative base pairing. It is not clear if these signals are significant, but such interaction may be important in restoring cooperative behavior to the *FN* riboswitch. It is unclear whether these upstream sequence elements are actually present in the riboswitch; there is so far little work in mapping the transcription start sites for natural riboswitches expressed in their native cellular environments.

4. <u>Aptamer 1–3´interaction (CCCC/GGGG).</u> Nearly all glycine riboswitches have a hairpin extension on the P3 stem of aptamer 1, although this is missing in the most widely studied riboswitches from *F. nucleatum* and *V. cholerae*. Further, nearly all of these P3b stems display a CCCC sequence in their apical loops (with occasional replacement of a C to U). Interestingly, these riboswitches also exhibit the sequence GGGG at their 3´ ends (immediately after P1 for aptamer II). Visual inspection of the available crystallographic models for the *FN* riboswitch and *VCII* dimer suggest that such sequences should be able to interact in the glycine-bound state of these riboswitches. Further evidence for this interaction comes from the observation that sequences without the 3´ GGGG sequence do not have the C-rich P3b loop in aptamer I. This interaction has presumably not been noticed before due its absence in the small glycine riboswitches chosen for study to date.

## Supporting References

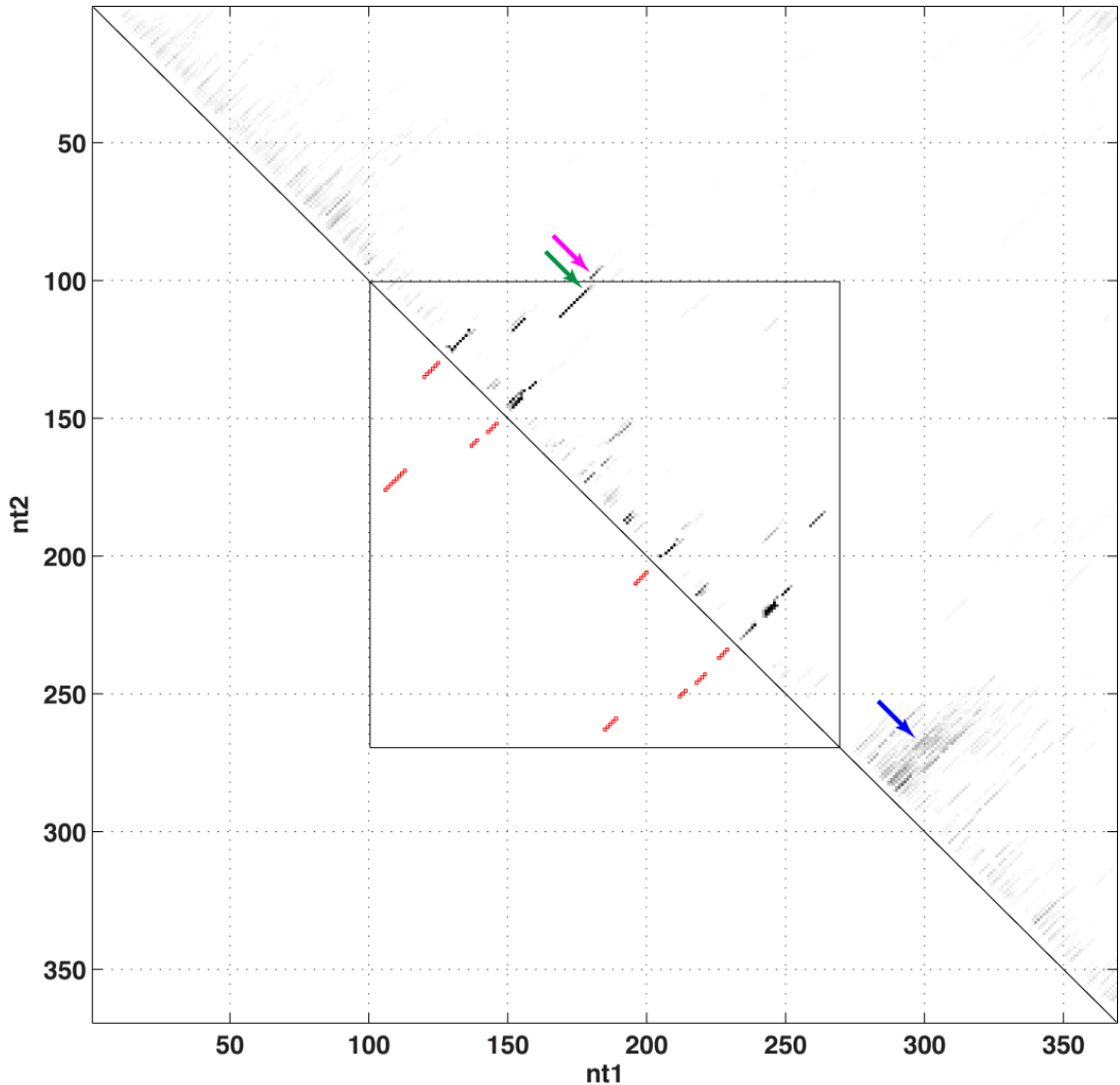

**Supporting Figure S1. Results of the BPPalign method on the glycine riboswitch sequences.** Grayscale gives average predicted Watson-Crick base pairing probability *bpp* across 360 glycine riboswitches, at positions that were aligned to the *FN* riboswitch sequence in the Rfam alignment. White represents pairs with average $bpp < 0.005$; black represents pairs with average $bpp \geq 0.1$. The large box gives the conventional bounds of the glycine-binding domains, e.g., as defined in Rfam or in previous studies[21,24]. Red squares give base pairs used to seed the Rfam alignment. Magenta arrow points to new predicted stem ("P0"); green arrow points to extension of P1. The 3-nt vertical spacing between these features represents a bulge of conserved length. Other features in the plot correspond to the assumed seed stems (c.f. red squares below diagonal and dark stripes across diagonal); 3′ stems involved in 'expression platforms' (translation repressor and transcription terminator stems; blue arrow); and likely false positives involving sequences conserved for other reasons (e.g., glycine binding) that happen to be complementary. The latter features are under further investigation.

```
Cluster:    1 -> model: CL_1.0     count:  183 occur_(%):  50.70 score:  10.142 bpp:  0.018 MI:  0.207 H:  0.208 cols:  227  357
Cluster:    2 -> model: CL_1.0     count:   52 occur_(%):  12.81 score:   9.232 bpp:  0.014 MI:  0.178 H:  0.207 cols:  578  356
Cluster:    3 -> model: CL_1.0     count:  192 occur_(%):  48.47 score:  10.062 bpp:  0.151 MI:  0.227 H:  0.241 cols:  583  703
Cluster:    4 -> model: TGA_1.0    count:   75 occur_(%):  20.06 score:  10.550 bpp:  0.294 MI:  0.279 H:  0.305 cols:   98  442
Cluster:    5 -> model: KT_1.0     count:  101 occur_(%):  28.13 score:  16.033 bpp:  0.183 MI:  0.201 H:  0.185 cols:  438   96
Cluster:    6 -> model: TGA_1.0    count:   62 occur_(%):  16.71 score:  11.553 bpp:  0.093 MI:  0.103 H:  0.086 cols:  422  443
Cluster:    7 -> model: TGA_1.0    count:   54 occur_(%):  15.04 score:  11.767 bpp:  0.086 MI:  0.033 H:  0.051 cols:  436  773
Cluster:    8 -> model: TGA_1.0    count:   32 occur_(%):   8.36 score:  11.028 bpp:  0.052 MI:  0.271 H:  0.464 cols:  442  466
```

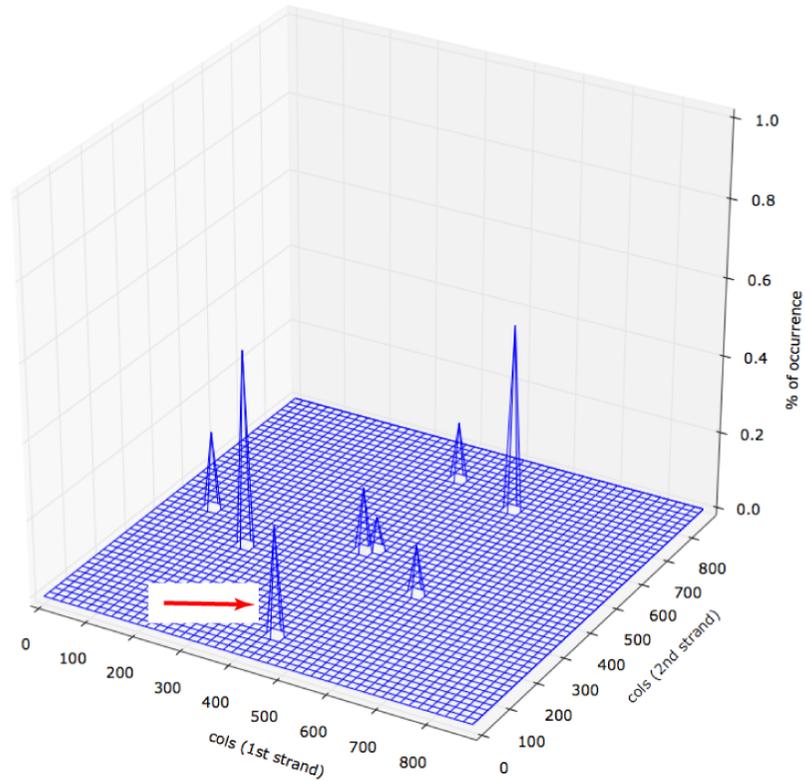

**Supporting Figure S2. Results of the *RMdetect* algorithm on sequence alignment of 360 glycine riboswitches**. Arrows point to kink-turn prediction. Additional tandem G/A ('TGA') predictions are partially correct or part of kink-turn purine-pairs motif. The C-loop motif ('CL') predictions appear to be false positives.

**12_cWW-tSS--tSH-tHS-tHS-cWW**

| Result id | Filename | Discrepancy | 1 | 2 | 3 | 4 | 5 | 6 | 7 | 8 | 9 | 10 | 11 |
|-----------|----------|-------------|---|---|---|---|---|---|---|---|---|----|----|
| Group_311 1 | IL_3PYO_043 | 0.0000 | C 1208 | G 1209 | A 1210 | G 1212 | A 1213 | A 1214 | G 1215 | U 1234 | G 1235 | G 1236 | A 123 |

Previous | Next
☑ Group_311 1

➕ Share

☐ Stereo on/off  ☐ nucleotide numbers on/off  ☐ 16A neighborhood    ☐ Show/hide all

**Supporting Figure S3. Results of the *JAR3D* algorithm on putative kink-turn element from the *FN* double glycine riboswitch**. A screen-shot of the top hit is shown.

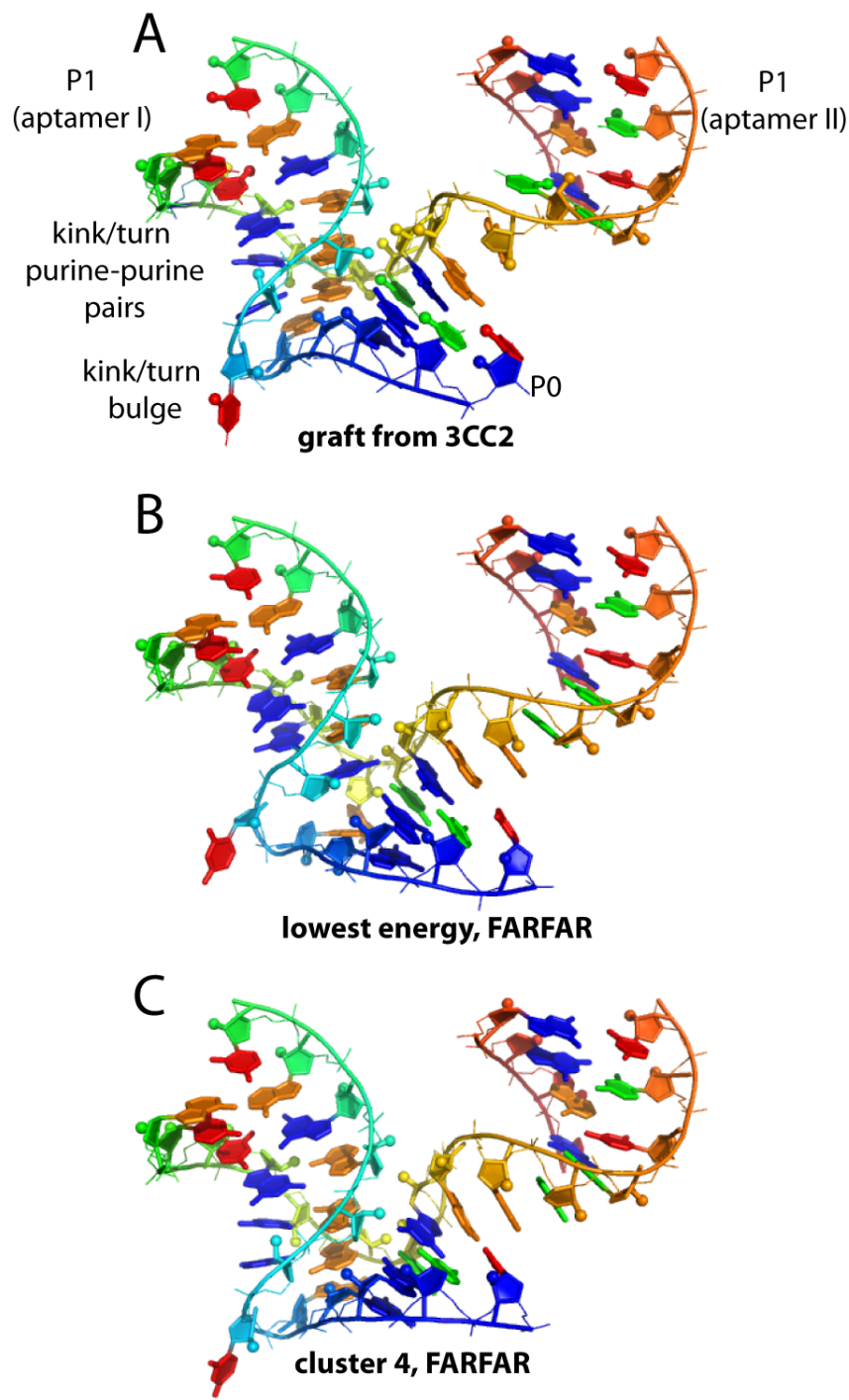

**Supporting Figure S4. Rosetta models for the new elements of the glycine riboswitch**. (A) Graft of the kink-turn from *H. marismortui* large ribosomal subunit crystallographic model (KT-46) into *FN* riboswitch model (PDB ID: 3P49), optimized with Rosetta. (B) Lowest energy model from *de novo* FARFAR modeling for the entire 8-nt linker strand and 9 added nucleotides; model recovers two purine-purine pairs in P1 and the novel three-Watson-Crick-base-pair stem P0. (C) Fourth lowest energy cluster from *de novo* FARFAR modeling fully recovers the complete kink-turn motif and P0. Bases are colored according to identity (A, orange; C, green; G, blue; U, red), and backbones are colored according to position in the motif. Riboswitch coordinates beyond the P1 stems of the double aptamers are not shown.

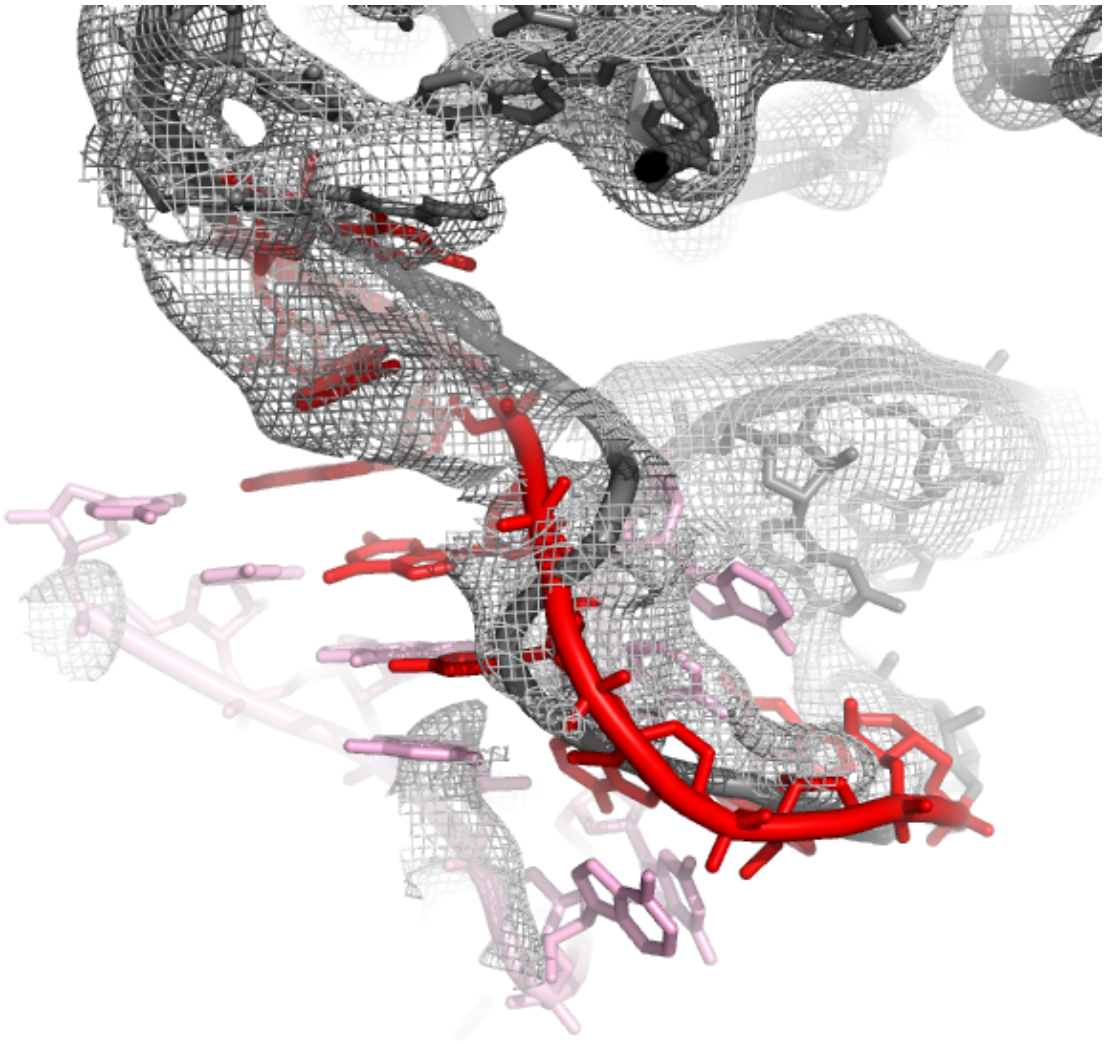

**Supporting Figure S5. Unexpected concordance of remodeled linker structured by the kink-turn motif and crystallographic coordinates of unstructured linker.** The Rosetta 3D model of the newly predicted inter-aptamer kink-turn (red and pink) reorganizes the inter-aptamer linker (red) through pairing with nine nucleotides (pink) 5′ of the conventional (minimal) glycine riboswitch boundary. The rebuilt linker path still approximately traces the path of the linker in the original glycine riboswitch coordinates (black) as well as the observed electron density (gray mesh).

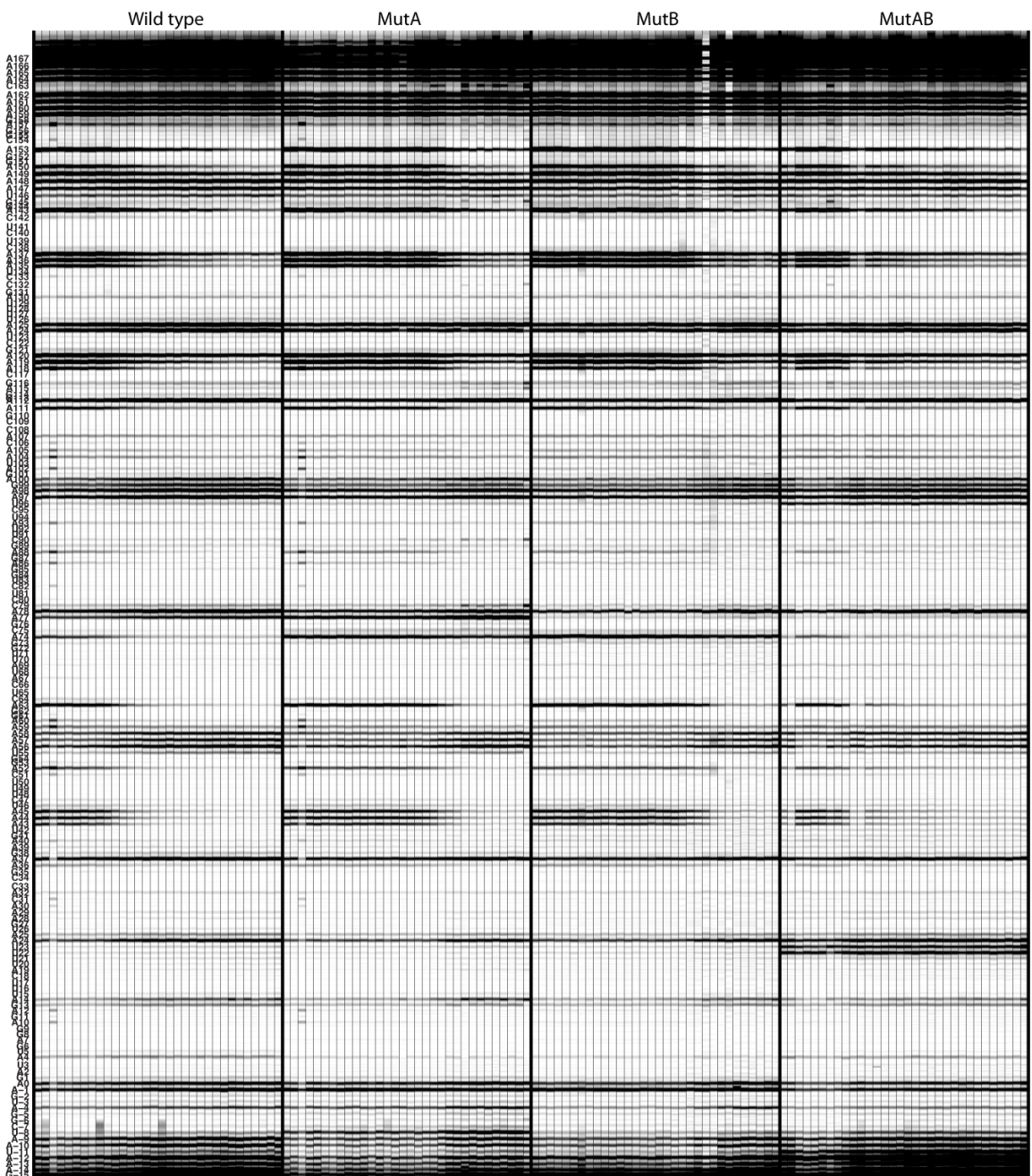

**Supporting Figure S6. DMS mapping data for glycine titrations of riboswitches and mutated variants.** Capillary electropherograms, aligned by HiTRACE, are shown for the wild type, MutA, MutB, and MutAB variants of the FN-KT RNA, probed at 24 °C in 0.5 mM MgCl₂, 50 mM Na-HEPES, pH 8.0. Within each set, glycine concentrations were (from left to right in μM): 0, 0.1, 0.2, 0.3, 0.5, 1, 1.5, 2, 3, 5, 10, 15, 20, 30, 50, 100, 150, 200, 300, 500, 1000, 1500, 2000, 3000, 5000, 10,000, 20,000, 30,000, 50,000, 100,000, 200,000, and 500,000. Minor bands from nuclease contaminants are visible in wild type and MutA (lane 3). One replicate titration for each mutant is shown, and correspond to the data given in main text Fig. 3. Additional replicates (total of two for wild type and MutA; total of three for MutB and MutAB) were carried out for each RNA variant and used to determine fitted equilibrium constants.

## A. State **0** to State **1**

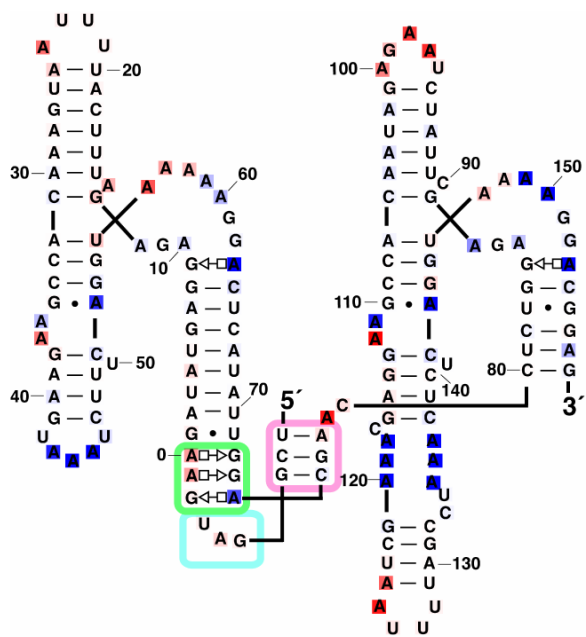

## B. State **1** to State **2**

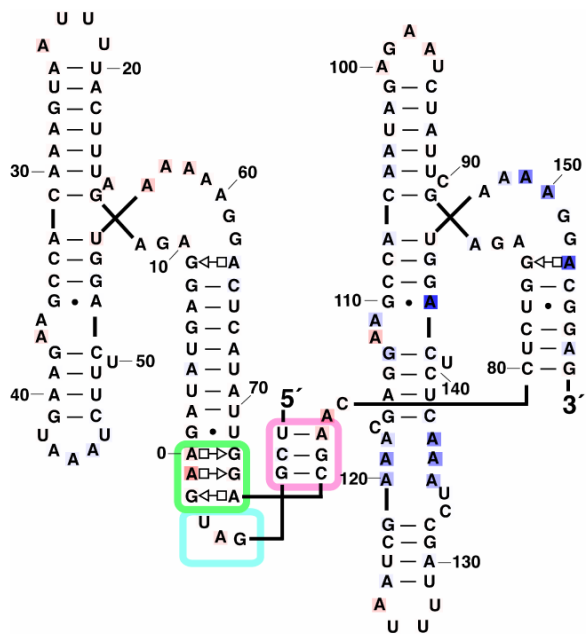

**Supporting Figure S7. DMS reactivity changes during glycine binding.** Coloring gives change in reactivity between glycine-free state and single-glycine-bound state (A) and between single-glycine-bound state and double-glycine-bound state (B). Blue and red represent protections and increasing reactivity, respectively. The second transition involves conformational transitions mainly in aptamer II. The simplest model then is that the first transition involves glycine binding to the aptamer I site and concomitant formation of inter-aptamer contacts; and the second transition involves glycine binding to the aptamer II site. Supporting this model, disruption of the kink-turn and P0 in the inter-aptamer linker gives a 100-fold shift in the glycine requirement for the concerted inter-domain change.